# The Statistical Software Revolution in Pharmaceutical Development: Challenges and Opportunities in Open Source

Daniel Sabanés Bové*,1, Heidi Seibold2, Anne-Laure Boulesteix3,4, Juliane Manitz5, Alessandro Gasparini6,7, Burak K. Günhan8, Oliver Boix9, Armin Schüler10, Sven Fillinger11, Sven Nahnsen11,14, Anna E. Jacob3,15, Thomas Jaki12,13
1 RCONIS / inferential.biostatistics GmbH, Friedrichstrasse 12, 4055 Basel, Switzerland
2 Digital Research Academy, Bayerstr. 77c, 80335 Munich, Germany
3 Institute for Medical Information Processing, Biometry and Epidemiology, Faculty of Medicine, LMU Munich, 81377 Munich, Germany
4 Munich Center for Machine Learning, Munich, Germany
5 R Validation Hub
6 Department of Medical Epidemiology and Biostatistics, Karolinska Institutet, PO Box 281, 17177 Stockholm, Sweden
7 Red Door Analytics AB, Valfrid Palmgrens gata 4, 11369, Stockholm, Sweden
8 Merck Healthcare KGaA, Frankfurter Strasse 250, 64293 Darmstadt, Germany
9 Bayer AG, Aprather Weg 18a, 42113 Wuppertal, Germany
10 MorphoSys GmbH, Semmelweisstr. 7, 82152 Planegg, Germany
11 Quantitative Biology Center (QBiC), University of Tübingen, 72076 Tübingen, Germany
12 Faculty of Informatics and Data Science, University of Regensburg, Bajuwarenstraße 4 93053 Regensburg, Germany
13 MRC Biostatistics Unit, University of Cambridge, East Forvie Building, Forvie Site, Robinson Way,
Cambridge CB2 0SR, UK
14 Biomedical Data Science, Department of Computer Science, University of Tübingen, 72076 Tübingen, Germany
15 National Food Institute, Technical University of Denmark, Anker Engelunds Vej 1, 2800 Lyngby, Denmark

## Abstract

Advances in technology and clinical trial designs have accelerated the need for new analytical and statistical methods. Open-source software development is increasingly the preferred solution for keeping pace with these technological advances in the pharmaceutical industry. However, with a long history of relying on licensed analysis software, there are philosophical and organizational barriers to overcome. In particular, the sustainability, reliability, useability, and feasibility of maintaining open-source statistical software long-term must be ensured. In this article, we describe the open-source revolution that is emerging in the pharmaceutical industry and how it facilitates greater scaling of innovative analytical methods in statistics. We discuss typical challenges to open-source software adoption and propose strategies to help overcome these challenges. Furthermore, we illustrate the potential for open-source software development to facilitate faster adoption and scaling of innovative analytical methods with powerful examples of successful projects, which highlight the roles of cross-company collaboration, career paths, education, and community building. Finally, we discuss the future of open-source and licensed statistical software.

# 1Introduction

The pharmaceutical industry has been described by many as having serious challenges with productivity, efficiency, and innovation (Pammolli et al 2011; Scannell et al 2012; Laermann-Nguyen & Backfisch 2021). Productivity, as assessed by the number of new drug approvals per year, has remained relatively flat while the research and development (R&D) spend per approval has increased dramatically (Schlander et al 2021). Various reasons have been posited including changes in portfolio composition (e.g. entering new high-risk therapeutic areas with lower success rates), greater complexity in some therapeutic areas due to greater competition, advances in technology leading to a wider array of potential targets, and the high degree of regulation governing the industry. A generation of leaders has responded to these challenges by investing in a myriad of new technologies. Examples include high-throughput screening, whole genome sequencing, and more recently, foundation models, machine learning, and artificial intelligence. These investments aim to accelerate target identification, personalise medicine, measure disease in new ways (e.g. smartwatches), predict outcomes, automate analysis and reporting, improve success rates, and use resources more efficiently.

To realise the value of investments in technology, careful consideration should be given to ensure that the tools available in R&D deliver value (Cook et al, 2014). New data modalities and endpoints, higher volumes of data, novel trial designs, and evolving regulations require innovation in the methods used to collect, analyse and report evidence about the safety and efficacy of new drugs (Viraswami-Appanna et al, 2024). Quantitative scientists also have significant roles in helping to fully realise the value of new technologies in drug development–gaining efficiencies through novel designs and devising new methods to fully assess patient benefit. Open-source software development has emerged as an agile solution for keeping pace with advances in statistical methodology. Recent methodological literature reveals that the standard no longer includes only code or data alongside a scientific publication, but also open-source software packages as reference companions (e.g. Sharma et al., 2024). As described by Rufibach et al (2024), there has been a conscious decision within pharmaceutical companies to develop open-source software for statistical innovations with the objective of increasing credibility and facilitating faster adoption in the statistical community.

This work examines the open-source statistical software revolution developing in the pharmaceutical industry, the philosophical and technical challenges associated with the adoption of open-source statistical software, and enabling strategies to address its sustainability and rapidly evolving future. We note that the challenges experienced in applied sciences and academia are very similar, and many of the solution strategies have parallels in both divisions. While we discuss academic examples, particularly in sections 3.3 and 4.3, we keep the focus on statistical analyses in the pharmaceutical industry to manage the scope of our discussion. We highlight the potential of open-source statistical software development with relevant real-life examples.

# 2 Organisational Challenges

## 2.1 The Elephant in the Room: Why Should we Bother?

The pharmaceutical industry has long relied on licensed closed-source proprietary statistical software as a secure choice. These software types offer documented validation and certified quality, which is included in the licensing fees. However, with rapid innovation in data and methods in the pharmaceutical industry, software requirements are changing, and pharmaceutical companies must identify their system requirements for the future. The product offering of vendors may no longer align with a company's rapidly developing needs. The company then has a choice: work with the vendor to try to make the product fit or develop an open solution for the future.

One of the most fundamental challenges senior executives pose against developing open-source statistical software in-house is the idea that it is not their core business. As shown in Figure 1, a pharmaceutical executive might wonder:

> *We are in the business of developing medical products to treat patients.*
> *So why should we get into internal statistical software development?*

It is a logical question and the response is just as rational, providing a counter example:

> *We also have accountants and lawyers.*
> *Although we are not in the business of banking or law, navigating those elements is essential to a successful business.*

In reality, the amount of data in big data has been growing for decades. New digital and imaging measurement modalities generate gigabytes of data per patient. We can integrate data from external sources to inform decisions, and in some cases supplement clinical trials. We are also implementing more complex clinical trial designs, allowing for adaptation throughout the process, even in confirmatory trials. All of this requires the use of increasingly sophisticated analysis methods. To implement these now and at scale, analytical method development needs to be accompanied with validated software that is re-usable, customisable, and flexible enough to handle the nuances of the relevant context. An ad-hoc approach to these challenges has its limitations, and we must recognise the need for dedicated Research Software Engineering (RSE) expertise, just as other expertise are valued in other business domains.

Hesitancy to commit to a coding language in an evolving space is another potential hurdle for corporate adoption of open-source software. A current debate revolves around the question: why should we use R, and not Python, or whatever will come next? Python may be a better general programming language and is preferred for machine learning and AI (Loukides, Martino 2020). For statistical analyses required for clinical trials, including hybrid datasets (which combine clinical trials with real world data), R is still advantageous due to the availability of statistical packages and wide adoption in academia, where new statistical methods are often developed. Importantly, large language models are rapidly emerging as a potential tool to simplify the translation of code between programming languages. In addition, we must acknowledge that both R and Python have been established programming languages in data science for more than two decades. It would be a misconception to think that predominant programming languages in a given application field change every few years. As a result, there is potentially greater risk and financial cost associated with remaining stagnant rather than adapting to the advancements and forward shift of the field.

## 2.2 Barriers and Silos within the Company

Rigid policies and regulation within the pharmaceutical industry are frequently cited as barriers to implementation of innovation in this space (Scannell et al, 2012). Variations in the interpretation of regulatory requirements (such as the Title 21 CFR Part 11 regulations on electronic records and signatures), can hinder the adoption of open-source statistical analysis software by companies (Schwartz et al, 2018). These regulations exist to ensure that the data and the systems in which they are operated on and stored are of sufficient reliability and quality; the corresponding requirements are defined on a high abstraction level. While health authorities like the United States Food and Drug Administration (US FDA) and European Medicines Agency (EMA) do not mandate specific software, they emphasise the need for thorough documentation of software packages used in submissions, including version and build identification (US FDA, 2015). Similarly, the ICH E9 guidelines (1999) highlight the importance of analysis software validity and reliability for the credibility of numerical results, requiring documentation of testing procedures.

In early 2020, the R Validation Hub, a collaboration to support the adoption of R within a biopharmaceutical regulatory setting, published a white paper introducing "A risk-based approach for assessing R package accuracy within a validated infrastructure" (Nicholls et al., 2020). The framework addresses concerns raised by statisticians, statistical programmers, informatics teams, executive leadership, quality assurance teams, and others within the pharmaceutical industry about the use of R and selected R packages as a primary tool for statistical analysis for regulatory submission work. The overall conclusion was that there is minimal risk in using base R and recommended packages as a component in a validated system for regulatory analysis and reporting (The R Foundation for Statistical Computing, 2021).

Beyond regulations, the presence of silos (defined by the Cambridge Dictionary as parts of a company, organization, or system that do not communicate with, understand, or work well with other parts) within companies can also affect the adoption and performance of open-source software (de Waal et al., 2019). Silos create inefficiencies in several notable ways:

1. Departmental Silos: Communication gaps between departments like IT and Statistics hinder the adoption of open-source software. Statistics departments need IT support for infrastructure (laptops, servers, and cloud-based computing facilities). Lack of communication can lead to frustration and prevent the Statistics department from accessing necessary modern tools.
2. Purpose Silos: Different systems and programming languages are often used for regulatory submissions versus exploratory statistical analyses. This requires recoding when exploratory analyses become relevant for regulatory use, causing time and resource inefficiencies. The work of the «Comparing Analysis Method Implementations in Software» (CAMIS) working group, which tries to align the results of different programming languages, shows that this is a non-trivial issue (Jen et al, 2023).
3. Project Silos: Within a company or institution, teams may copy and paste code between projects instead of developing generic, reusable code. This leads to manual adaptation, potential errors, and a lack of resource efficiency, especially as code bases grow more complex over time. Strategic planning for joint development efforts could prevent this.

## 2.3 Overcoming Proprietary Mindsets

In the competitive field of pharmaceutical development, where the aforementioned in-house lawyers are protecting intellectual property, organizations may take for granted the need to keep statistical software and code development proprietary. With rather established industry data, reporting, and output standards, companies are well-poised to gain efficiencies by using common open-source statistical software and pipelines for standard outputs. For new innovations in statistical analyses, the sharing of code can quickly facilitate adoption (Sojer, Henkel, 2010) –thereby normalising new methods for regulators and other stakeholders.

The repeated effort necessitated to keep statistical coding and analysis pipelines proprietary is redundant. Standard statistical analyses are not the subject of innovation, research, or the business itself, but only used as tools to present the evidence generated, e.g. from clinical trials. The intellectual property is the medicine itself. Pharmaceutical companies employ qualified personnel to perform their respective roles, but across companies, the produced data, analysis, and outputs are similar. Efficiency gains may be realised by collaborating and co-developing methods and statistical software. Expediting statistical software availability ensures that the most profitable and beneficial aspects of drug development, the new medical products, can be delivered sooner to patients while being simultaneously more cost efficient (see also Strauss 2010).

# 3. Success Factors for Open-Source Statistical Software

Statistical software is an invaluable tool in quantitative decision making and scientific research. In pharmaceutical companies' top management, emphasis is often placed on the ease of visualizing and interpreting results to inform decision making, assuming that the analyses and resulting numbers are correct. However, behind the scenes, multiple steps must be taken to ensure the usability and quality of statistical software.

## 3.1 Usability

Open-source projects often start from code written by an individual for their own use (e.g. to explore a novel statistical methodology) which is eventually published either due to altruistic motives or external pressures and incentives. However, others may sometimes struggle to use the software due to lack of suitable documentation, intuitive interface, or naming conventions. This results in researchers sometimes preferring to start from the beginning instead of using the already available software. Simultaneously, the usability of the software has a profound influence on its subsequent success.
Application Program Interface (API) design itself can greatly enhance the usability of software libraries (Myers and Stylos, 2016). Even simple elements such as the semantics of method signatures (i.e. the names of the methods, the parameters or inputs and their order) can make a difference. Usability tests can help in designing efficient and effective user interfaces or client APIs and should be a prioritized part of the development process.

## 3.2 Quality: Accuracy, Reliability, and Reproducibility by Design

The accumulation of scientific knowledge fundamentally relies on the quality and reproducibility of analysis results (Peng et al., 2006; Begley & Ioannidis, 2015; Stupple et al., 2019). Manual analysis workflows and non-version-controlled code development are major obstacles to achieving reproducibility. To ensure accurate analyses and results, all new software development should incorporate professional workflows and adequate testing.  Sanchez et al. (2021) described key steps for implementing a code quality assurance process that researchers can follow throughout a project to improve their coding practices regarding the quality of the final data, code, analyses, and results. This includes code style, documentation, version control, data management, testing, and code review. Taschuk and Wilson (2017) presented ten simple rules to make research software robust enough to be run reproducibly. The rules included version control, documentation, release versioning, build tools, and tests, among others.

To support the development and use of new R packages, risk assessment frameworks and tools have been developed (Manitz et al, 2022). For practical purposes, risk assessment criteria may be categorised into package purpose, good maintenance practices, testing coverage, and community usage. The R Validation Hub also created tools like the riskmetric R Package and the Risk Assessment application to simplify risk assessment information gathering (R Validation Hub et al., n.d., 2020/2022). In addition, the Regulatory R Package Repository working group was established (RC Working Group on Repositories, 2021/2022) and aims to create a repository of R packages which are deemed reliable enough to be used for medical data analysis. The idea is to help users differentiate easily between high-quality and standard packages and thus make the most informed decisions when selecting software for a given purpose. The development of statistical software —from the first prototype to high-quality reliable packages — is indeed a long-term process that can be conceptualised in distinct “phases”, analogous to the phases of drug development (Heinze et al., 2024). It is essential to help software users determine the development phase a piece of software has reached, enabling them to make informed decisions.

Considering that no method is expected to perform best across all settings, method selection can also be informed by simulation studies tailored to the specific context under consideration. Implementing such simulation-based comparison studies in an unbiased way requires advanced skills and major efforts (Boulesteix et al., 2024). Open-source software that implements such complex studies can, in the long run, contribute to better informed methodological decisions. See Ruberg et al. (2024) and Wahab et al. (2024) for recent examples addressing subgroup identification methodologies and methods for drawing pertinent causal inferences in trials with intercurrent events, respectively.

## 3.3 Collaboration for Efficiency and Reliability

For both industry and academia, publication of open-source statistical code is important. External stakeholders of pharmaceutical companies, such as the regulatory authorities, health insurance payer institutions, legislators, and the public, rightfully want to be informed on how the data from clinical trials was analysed and summarised into the final published results. Here. the possibility of sharing clinical trial data is a first step (Hopkins et al., 2018), but the availability of the full stack of statistical software used in the data analysis is an important second step.

In academia, the availability of software used to implement new statistical methods in online appendices is an increasingly common requirement for publishing. With the publication of open-source software, the next logical step is for cross-organisational collaboration on this common public code base. Additionally, the software becomes more reliable when the burden of developing, testing, and addressing user requests is shared across more developers.

Within the pharmaceutical industry, several initiatives drive the synergistic collaboration between companies on common open-source statistical software.

- The R consortium provides support to the R Foundation and to the key organizations developing, maintaining, distributing, and using R software through the identification, development, and implementation of infrastructure projects (R: The R Foundation, n.d.; RConsortium, n.d.). It hosts several working groups that work on specific topics, e.g. submissions, tabulations, certifications, and repositories.
- The openstatsware working group aims to engineer selected R packages to fill gaps in the open-source statistical software landscape and to promote good software engineering practices within biostatistics (ASA Biopharmaceutical Section Software Engineering Working Group, 2022).
- The PHUSE organization is a global community and platform for the discussion and progress of statistical programming topics, and
- PSI is a community dedicated to leading and promoting the use of statistics within the healthcare industry (PSI Web, 2018; Warren, 2022).

# 4. Success Strategies: Case Studies and the Power of Collaborative Ecosystems

## 4.1 Insights from Recent Case Studies

Table 1 provides examples of five open-source software packages developed to implement statistical innovations in response to evolving business needs, as well as increasing the efficiency of established statistical reporting. The first example, rbmi implements a computationally efficient, fully deterministic approach to imputing missing data in clinical trials with longitudinal endpoints that controls type I error and is consistent with the research question (Gower-Page et al 2022; Wolbers et al 2022). The second example, crmPack, implements a wide range of model-based dose escalation designs, ranging from classical and modern continual reassessment methods (CRMs), based on dose-limiting toxicity endpoints and dual-endpoint designs (e.g., plus biomarker or clinical endpoints), and permitting Bayesian and non-Bayesian inference (Sabanés Bové et al 2019; Boix & Guenhan, 2022). The third example, bonsaiforest, aims to more accurately obtain treatment effect estimators for subgroups in the context of binary and time to event endpoints (Vazquez Rabunal et al, 2024). The fourth example, tern, is a Table, Listings, and Graphs (TLG) library for common outputs used in clinical trials (Zhu et al 2025). It is merely one example of how new software can help to make statistical reporting more efficient, moreover enabling open-source based submissions to regulatory authorities. The final example, rpact, is a fully validated package that enables the design, simulation, and implementation of fixed and confirmatory clinical trial designs (Wassmer & Pahlke, 2025). Each of these examples solves very relevant analysis challenges in drug development using rigorous, validated methods that have been published in peer reviewed journals. The fact that the software is publicly available helps to ensure that the community can adopt it quickly, identify deficiencies and errors, contribute ideas and additions, and resolve issues promptly.

These case studies share a common factor—a collaborative ecosystem for quality and sustainability. Here, we deem a software project sustainable if the resulting software can be used over a prolonged period, thus eventually saving more time with its use compared to the time it took to develop the software overall. Although these case study software projects were initially developed under the sponsorship of a primary backer, their continued maintenance and evolution are now made possible through the pooled resources of multiple sponsors and the active participation of a wide user community. This collaborative model ensures that the software remains relevant, up-to-date, and able to meet the evolving needs of the pharmaceutical industry. Additionally, it fosters a sense of shared ownership and responsibility among the users, who are invested in the software's success and are more likely to contribute to its ongoing development.

## 4.2 Building Sustainable Teams for the Future

Analysis methods are increasing in their complexity, and the volume of data in general is growing rapidly. Implementing Research Software Engineering (RSE) as a recognized and dedicated profession, jointly with a basic RSE skills education for all statisticians, would provide the special expertise needed to link both the complexity of statistical analysis and burden of data volume (e.g. Bannert, 2024). Research Software Engineering is the use of software engineering principles in research projects to create reproducible, reusable, reliable, and accurate software. It requires an understanding of both the research subject and software development to ensure software can effectively support research outcomes (Woolston, 2022). This needs to be based on the honest recognition that software engineering is more than just being able to code in a programming language – in particular, most statisticians do not have software engineering skills. Research software engineers are crucial collaborators with scientists of other professions, and the RSE collaborations would not only include multidisciplinary collaboration within an organization, but also structured cooperation across both academia and industry.

Given their expertise in research, statistics, and software engineering, RSEs are highly sought-after by both academia and industry. To attract and retain these valuable talents, organizations must cultivate an environment that recognizes and rewards their contributions. Within academia, this begins with acknowledging software as a valuable research output and incentivising its development (Kuzak et al., 2018; Taschuk & Wilson, 2017). While publications remain a cornerstone of academic recognition, a growing movement advocates for software to be considered a first-class research output, granting due credit to contributors and maintainers. Creating dedicated career paths for RSEs that extend beyond postdoctoral positions, including leadership roles is needed, similar to those of librarians and lecturers (e.g. as implemented by the Quantitative Biology Center at the University of Tübingen, or the National Coordination Point Research Data Management in the Netherlands, see LCRDM, 2025).

Likewise, the industry demand for software engineers proficient in data science languages like R and Python is intensifying (Varney, 2018), extending beyond statistics to fields such as chemistry (Python Success Stories, n.d.). Competition for these highly sought-after individuals is fierce, with tech companies also vying for their skills. Consequently, the recruitment process can be resource-intensive, making the retention of RSEs through competitive career paths and compensation crucial. It is essential to recognize that RSEs are not merely support staff but integral members of the research ecosystem. Providing them with competitive salaries, seniority levels, and opportunities for growth comparable to those of methodology experts or managers is essential to maintain a motivated and productive workforce.

# 5. Discussion

In the highly regulated pharmaceutical development industry, the move to open-source statistical software is nothing short of revolutionary. At the time of this writing, we are in the middle of this trajectory as many large pharmaceutical companies make this transition. New open-source statistical software packages are emerging regularly in parallel with the analytical methods. Already, registration packages have been submitted to health authorities (across FDA, EMA, NMPA and PDMA) with analyses generated entirely using open-source statistical software (Patel & Chen, 2024, Knoph & Larsen, 2024).

Multiple factors have brought the industry to this point, including advances in the technology of other disciplines, newer data modalities, and more complex study designs driving the development of new analytical methods. After decades of relying almost exclusively on licensed and closed-source, proprietary statistical software, a key goal is to practise statistical software engineering in a sustainable way. This is accomplished by publishing methods and software as open source, integrating new software in an established ecosystem, organizing long-term maintenance, and adhering to good software engineering practices. This requires overcoming certain organizational mindsets and silos (within and across companies) as well as ensuring that the right team members (including those with necessary RSE skills) are hired and retained to develop and maintain software, while also cross-training colleagues to facilitate this work. To obtain the full potential of open-source software development, it is essential that organisations collaborate more effectively when developing software across both industry and academia settings, ultimately leading to an improved and successful product for all. Here the industry can learn from academia, which has relevant experience with RSE groups and roles, as well as cross-institutional cooperation.

The goal of this manuscript is not to imply that one must take an either/or decision on proprietary closed-source versus open-source statistical software, but rather we hoped to highlight that Pandora's box has already been opened, and that open-source statistical software can assist industry and academia with large demands to innovate for the future.

# Acknowledgments

Thomas Jaki received funding from the UK Medical Research Council (MC_UU_00040/03). For the purpose of open access, the author has applied a Creative Commons Attribution (CC BY) license to any Author Accepted Manuscript version arising.

# References

ASA Biopharmaceutical Section Software Engineering Working Group. (2022). ASA BIOP SWE WG. https://openstatsware.org

Bannert, Matthias. (2024). Research Software Engineering: A Guide to the Open Source Ecosystem. https://doi.org/10.1201/9781003286899.

Begley, C. G., & Ioannidis, J. P. A. (2015). Reproducibility in Science. Circulation Research, 116(1), 116–126. https://doi.org/10.1161/CIRCRESAHA.114.303819

Boix, O., & Günhan, B. K. (2022, August 23). A collaborative approach to software development; The crmPack experience. ISCB 2022 Conference, Newcastle, UK. https://www.burakguenhan.com/talk/crmpack/

Boulesteix, A. L., Baillie, M., Edelmann, D., Held, L., Morris, T. P., & Sauerbrei, W. (2024). Editorial for the special collection “Towards neutral comparison studies in methodological research”. Biometrical Journal, 66(2), 2400031.

Cook, D., Brown, D., Alexander, R. et al. Lessons learned from the fate of AstraZeneca's drug pipeline: a five-dimensional framework. Nat Rev Drug Discov 13, 419–431 (2014). https://doi.org/10.1038/nrd4309

de Waal, Weaver, Day, & van der Heijden. (2019). Silo-Busting: Overcoming the Greatest Threat to Organizational Performance. Sustainability, 11(23), 6860. https://doi.org/10.3390/su11236860

European Medicines Agency. (1998). ICH Topic E9 Statistical Principles for Clinical Trials (p. 37) [CPMP/ICH/363/96].

Gower-Page, C., Noci, A., and Marcel, M. (2022), “rbmi: AR Package for Standard and Reference-based Multiple Imputation Methods,” Journal of Open Source Software, 7, 4251. [4,5]

Heinze, G., Boulesteix, A. L., Kammer, M., Morris, T. P., White, I. R., & Simulation Panel of the STRATOS Initiative. (2024). Phases of methodological research in biostatistics—building the evidence base for new methods. Biometrical Journal, 66(1), 2200222.

Hopkins, A. M., Rowland, A., & Sorich, M. J. (2018). Data sharing from pharmaceutical industry sponsored clinical studies: Audit of data availability. BMC Medicine, 16(1), 165. https://doi.org/10.1186/s12916-018-1154-z

ICH E9 Expert Working Group. Statistical Principles for Clinical Trials: ICH Harmonized Tripartite Guideline. Statistics in Medicine 1999; 18: 1905–1942.

Min-Hua Jen, Brian Varney, Kyle Lee, Benjamin Arancibia, Mia Qi, Lyn Taylor, Christina Fillmore, Joseph Rickert, Mike Stackhouse, Michael Rimler (2023). Key Considerations When Understanding Differences in Statistical Methodology Implementations Across Programming Languages – An Introduction to the CAMIS Project. https://phuse.s3.eu-central-1.amazonaws.com/Deliverables/Data+Visualisation+%26+Open+Source+Technology/WP077.pdf

Knoph, A.S., Larsen, S.F. (2024). Our R-Based Submissions: A 1-year follow-up, Lessons Learned, and Future Strategies. R/Pharma 2024 abstract. https://rinpharma.com/publication/rinpharma_483/

Laermann-Nguyen, U., Backfisch, M. Innovation crisis in the pharmaceutical industry? A survey. SN Bus Econ 1, 164 (2021). https://doi.org/10.1007/s43546-021-00163-5.

LCRDM. (2025). Professionalizing the role of Research Software Engineers in the Netherlands. Zenodo. https://doi.org/10.5281/zenodo.15019998

Loukides, M., & Martino, C. (2020). *The State of Data Science 2020*. O'Reilly Media. URL: https://www.oreilly.com/radar/the-state-of-data-science-2020/

Kuzak, M., Cruz, M., Thiel, C., Sufi, S., & Eisty, N. (2018, November 28). Making Software a First-Class Citizen in Research. WSSSPE6.1 Speed Blog. https://software.ac.uk/blog/2018-11-28-making-software-first-class-citizen-research

Manitz, J., Nicholls, A., Gotti, M., Kelkhoff, D., Clark, A., Palukuru, U. P., & Taylor, L. (2022). Risk Assessment of R Packages: Learning and Reflections (No. 3; Biopharmaceutical Report, pp. 3–10). American Statistical Association. https://higherlogicdownload.s3.amazonaws.com/AMSTAT/fa4dd52c-8429-41d0-abdf-0011047bfa19/UploadedImages/BIOP%20Report/BioPharm_fall2022FINAL.pdf

Myers, B.A., Stylos, J. (2016). Improving API usability. Communications of the ACM 59, 6 (2016), 62–69. https://doi.org/10.1145/2896587

Nicholls, A., Bargo, P., R., & Sims, J. (2020). A Risk-based Approach for Assessing R package Accuracy within a Validated Infrastructure. https://www.pharmar.org/white-paper/

Pammolli, F., Magazzini, L. & Riccaboni, M. The productivity crisis in pharmaceutical R&D. Nature Rev. Drug Discov. 10, 428–438 (2011).

Patel H & Chen J. (2024). The R Journey to Submission, Retrieved 16 April 2025 from: https://phuse.s3.eu-central-1.amazonaws.com/Archive/2024/Connect/EU/Strasbourg/PAP_SA02.pdf.

Peng, R. D., Dominici, F., & Zeger, S. L. (2006). Reproducible Epidemiologic Research. American Journal of Epidemiology, 163(9), 783–789. https://doi.org/10.1093/aje/kwj093

PSI Web. (2018). PSI. https://www.psiweb.org

Python Success Stories. (n.d.). Python.Org. Retrieved November 30, 2022, from https://www.python.org/about/success/astra/

R: The R Foundation. (n.d.). Retrieved November 30, 2022, from https://www.r-project.org/foundation/

R Validation Hub, Gotti, M., Clark, A., Krajcik, R., Gans, M., Kallem, A., & Fission Labs India. (2022). PharmaR/risk_assessment [R]. pharmaR. https://github.com/pharmaR/risk_assessment (Original work published 2020)

R Validation Hub, Kelkhoff, D., Gotti, M., Miller, E., K, K., Zhang, Y., Miliman, E., & Manitz, J. (n.d.). Riskmetric [R]. Retrieved November 30, 2022, from https://pharmar.github.io/riskmetric/

RC Working Group on Repositories. (2022, November 14). https://github.com/RConsortium/r-repositories-wg (Original work published 2021)

RConsortium. (n.d.). R Consortium. Retrieved November 30, 2022, from https://www.r-consortium.org/

Ruberg, S., Zhang, Y., Showalter, H., & Shen, L. (2024). A platform for comparing subgroup identification methodologies. Biometrical Journal, 66(1), 2200164.

Rufibach, K., Wolbers, M., Devenport, J., Yung, G., Harbron, C., Bedding, A., … Wang, J. (2024). Implementation of Statistical Innovation in a Pharmaceutical Company. Statistics in Biopharmaceutical Research, 17(1), 113–124. https://doi.org/10.1080/19466315.2024.2327291

Sabanés Bové, D., Yeung, W. Y., Palermo, G., & Jaki, T. (2019). Model-Based Dose Escalation Designs in R with crmPack. Journal of Statistical Software, 89, 1–22. https://doi.org/10.18637/jss.v089.i10

Sanchez, R., Griffin, B. A., Pane, J., & McCaffrey, D. F. (2021). Best practices in statistical computing. Statistics in Medicine, 40(27), 6057–6068. https://doi.org/10.1002/sim.9169

Scannell, J., Blanckley, A., Boldon, H. et al. Diagnosing the decline in pharmaceutical R&D efficiency. Nat Rev Drug Discov 11, 191–200 (2012). https://doi.org/10.1038/nrd3681

Schlander, M., Hernandez-Villafuerte, K., Cheng, C.Y., Mestre-Ferrandiz, J., Baumann, M. How Much Does It Cost to Research and Develop a New Drug? A Systematic Review and Assessment. Pharmacoeconomics 39, 1243-1269 (2021). https://doi.org/10.1007/s40273-021-01065-y

Sharma NK, Ayyala R, Deshpande D, et al. Analytical code sharing practices in biomedical research. PeerJ Computer Science 28, 10:e2066 (2024). https://doi.org/10.7717/peerj-cs.2066

Schwartz, M., Harrel, F., Rossini, A., & Francis, I. (2018, March 25). R: Regulatory Compliance and Validation Issues A Guidance Document for the Use of R in Regulated Clinical Trial Environments. Retrieved from https://www.rproject.org/doc/R-FDA.pdf

Seibold, H., Charlton, A., Boulesteix, A., & Hoffmann, S. (2021). Statisticians, roll up your sleeves! There's a crisis to be solved. Significance, 18(4), 42–44. https://doi.org/10.1111/1740-9713.01554

Sojer, M. and Henkel, J. (2010). "Code Reuse in Open Source Software Development: Quantitative Evidence, Drivers, and Impediments," Journal of the Association for Information Systems, 11(12), https://aisel.aisnet.org/jais/vol11/iss12/2

Strauss, S. (2010). Pharma embraces open source models. Nat Biotechnol 28, 631–633 (2010). https://doi.org/10.1038/nbt0710-631

Stupple, A., Singerman, D., & Celi, L. A. (2019). The reproducibility crisis in the age of digital medicine. Npj Digital Medicine, 2(1), 2. https://doi.org/10.1038/s41746-019-0079-z

Szymański, R. (2022, December 10). CRAN and the Isoband Incident—Is Your Project at Risk and How to Fix It. https://appsilon.com/cran-and-the-isoband-incident/

Taschuk, M., & Wilson, G. (2017). Ten simple rules for making research software more robust. PLOS Computational Biology, 13(4), e1005412. https://doi.org/10.1371/journal.pcbi.1005412

The R Foundation for Statistical Computing. (2021). R: Regulatory Compliance and Validation Issues A Guidance Document for the Use of R in Regulated Clinical Trial Environments. https://www.r-project.org/doc/R-FDA.pdf

U.S Food & Drug Administration. (2015, May 6). Statistical Software Clarifying Statement. Retrieved from https://fda.report/media/109552/Statistical-Software-Clarifying-Statement-PDF.pdf

Varney, B. (2018, February 18). Why R? The Next Generation in Pharma. Pubs - Bio-IT World. https://www.bio-itworld.com/news/2022/02/18/why-r-the-next-generation-in-pharma

Vazquez Rabunal, M., Sabanés Bové, D., Wolbers, M. (2024). bonsaiforest: Shrinkage Based Forest Plots. R package version 0.1.1, https://CRAN.R-project.org/package=bonsaiforest

Viraswami-Appanna K, Buenconsejo J, Baidoo C, Chan I, Li D, Micsinai-Balan M, Tiwari R, Yang L, Sethuraman V. Accelerating drug development at Bristol Myers Squibb through innovation. Drug Discov Today. 2024 May;29(5):103952. doi: 10.1016/j.drudis.2024.103952. Epub 2024 Mar 18. PMID: 38508230.

Wahab, A. H. A., Qu, Y., Michiels, H., Luo, J., Zhuang, R., McDaniel, D., ... & Sabbaghi, A. (2024). CITIES: Clinical trials with intercurrent events simulator. Biometrical Journal, 66(1), 2200103.

Warren, K. (2022, October 2). Welcome to the PHUSE Advance Hub—WORKING GROUPS - PHUSE Advance Hub. PHUSE. https://advance.phuse.global/

Wassmer G, Pahlke F (2025). *rpact: Confirmatory Adaptive Clinical Trial Design and Analysis*. doi:10.32614/CRAN.package.rpact, R package version 4.2.0, https://cran.r-project.org/package=rpact.

Wolbers M, Noci A, Delmar P, Gower-Page C, Yiu S, Bartlett JW. Standard and reference-based conditional mean imputation. Pharm Stat. 2022 Nov;21(6):1246-1257. doi: 10.1002/pst.2234. Epub 2022 May 19. PMID: 35587109; PMCID: PMC9790242.

Woolston, Chris (2022). "Why Science Needs More Research Software Engineers: Ten Years After Their Profession Got Its Name, Research Software Engineers Seek to Swell Their Ranks". Nature. doi:10.1038/d41586-022-01516-2. PMID 35641618

Zhu J, Sabanés Bové D, Stoilova J, Garolini D, de la Rua E, Yogasekaram A, Wang H, Collin F, Waddell A, Rucki P, Liao C, Li J (2025). *tern: Create Common TLGs Used in Clinical Trials*. R package version 0.9.8.433, https://github.com/insightsengineering/tern.

Table 1. Case Studies: Open-Source Packages for Common Statistical Analyses in Drug Development

| Name | Application | Relevance |
|---|---|---|
| *rbmi* | Implements novel deterministic missing data imputation for trials with continuous longitudinal outcomes | Addresses missing data due to dropout or other intercurrent events in a computationally efficient manner that is more consistent with the research objectives than alternative methods |
| crmPack | Implements a wide range of model-based dose escalation designs. | Flexible package that implements standard and customized designs with a dose-limiting toxicity endpoint as well as dual endpoints (e.g., plus biomarker or clinical endpoints), and permitting Bayesian and non-Bayesian inference |
| bonsaiforest | Obtain more accurate treatment effect estimators for subgroups in the context of binary and time to event endpoints. | Interpreting subgroup-specific treatment effect estimates requires great care due to the smaller sample size of subgroups and the large number of investigated subgroups. This method borrows information across subgroups to appropriately shrink the subgroup estimate toward the population estimate. |
| tern | Create common tables, listings and graphs used in clinical trials. | Enables clinical trial analysis in R up to regulatory standards, in particular provides code for producing clinical trial outputs ready for submission to regulatory authorities. |
| rpact | Clinical trial planning, design, simulation, evaluation, and analysis | Fully validated free of charge R package with code, vignettes, an app, and training to enable design of fixed and adaptive clinical trials. |

Figure 1: Why do we need statistical software developers?

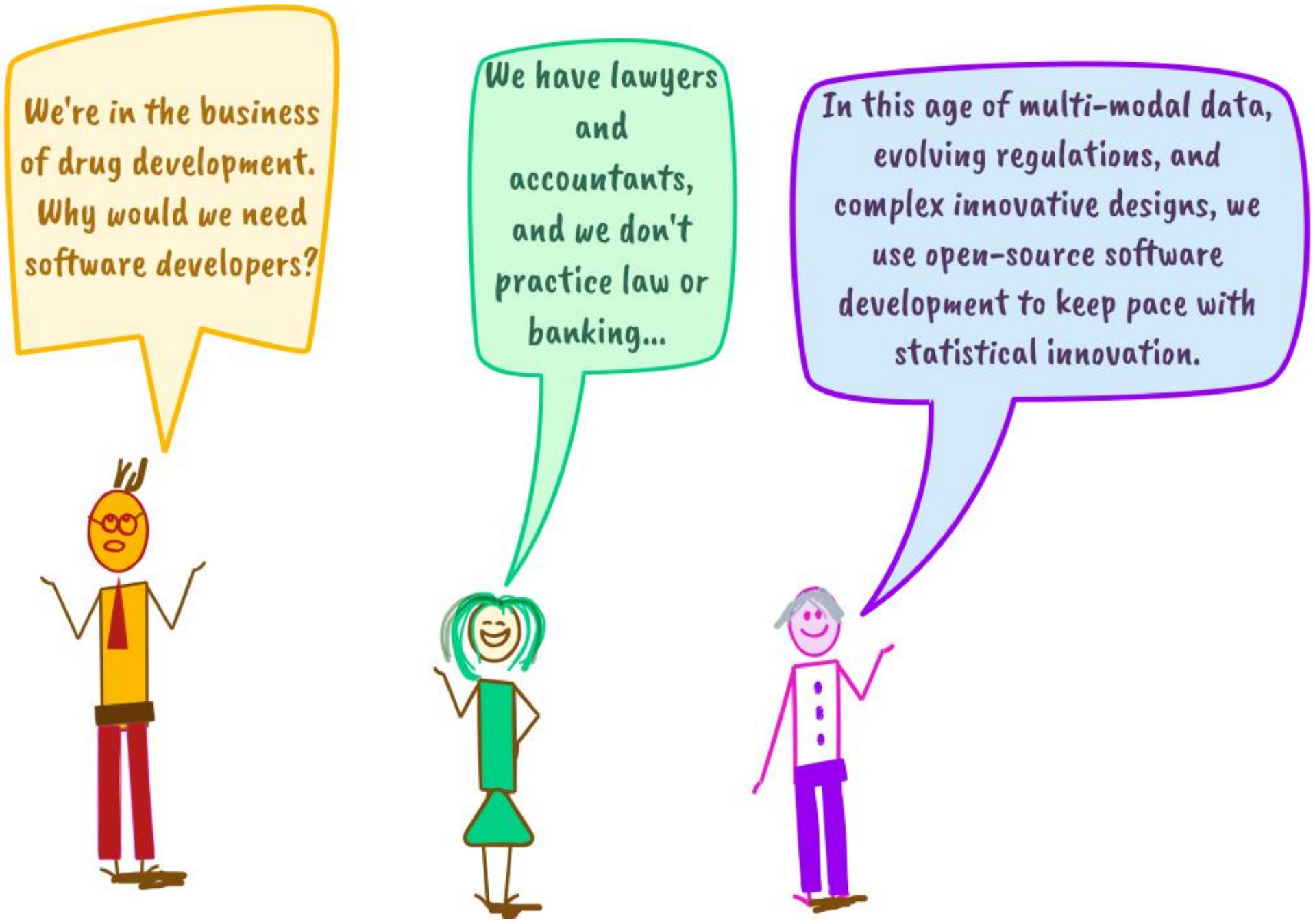